\documentclass[twoside,12pt]{article}%
\usepackage{epsfig}
\usepackage{amsmath}
\usepackage{amsfonts}
\usepackage{amssymb}
\usepackage{graphicx}%
\makeatletter

\def\vereq#1#2{\lower3pt\vbox{\baselineskip1.5pt \lineskip1.5pt
\ialign{$\m@th#1\hfill##\hfil$\crcr#2\crcr\sim\crcr}}}

\makeatother
\begin{document}

\title{\vspace{1cm} Radiative capture cross sections: challenges and solutions}
\author{C.A. Bertulani
\thanks{ E-mail: bertulani@nscl.msu.edu}\\
NSCL and Department of Physics and Astronomy, \\
Michigan State University, East Lansing, MI 48824\\
}
\date{}
\maketitle

\begin{abstract}
Radiative capture reactions are one of the main inputs in stellar modelling.
In numerous situations, the low energies at which these reactions occur in
stars are not accessible with present experimental techniques. Electron
screening is one of the major causes of problems in separating the bare cross
sections from the screened ones. Indirect experimental techniques have been
proposed and are now one of the main tools to obtain these cross sections. I
discuss the electron screening problem and the latest progresses obtained with
indirect methods.

\end{abstract}

KEYWORDS: Nuclear Astrophysics, Radioactive Beams.

PACS: 26.,25.60.-t

\nopagebreak

\section{Introduction}

\label{intro}

Present studies in nuclear astrophysics are focused on the opposite ends of
the energy scale for nuclear reactions: (a) the very high and (b) the very low
relative energies between the reacting nuclei. Projectiles with high
bombarding energies produce nuclear matter at high densities and temperatures.
This is the main goal at the RHIC accelerator at the Brookhaven National
Laboratory and also the main subject discussed in this Workshop.\ One expects
that matter produced in central nuclear collisions at RHIC for $\sim10^{4}$
GeV/nucleon of relative energy, and at the planned Large Hadron Collider at
CERN, will undergo a phase transition and produce a \textit{quark-gluon
plasma}. One can thus reproduce conditions existent in the first seconds of
the universe and also in the core of neutron stars. At the other end of the
energy scale are the low energy reactions of importance for stellar evolution
(see figure 1). A chain of nuclear reactions starting at $\sim10-100$ keV
leads to complicated phenomena like supernovae explosions or the energy
production in the stars.

\begin{figure}[ptb]
\begin{center}
\includegraphics[
height=3.in, width=6.5 in ] {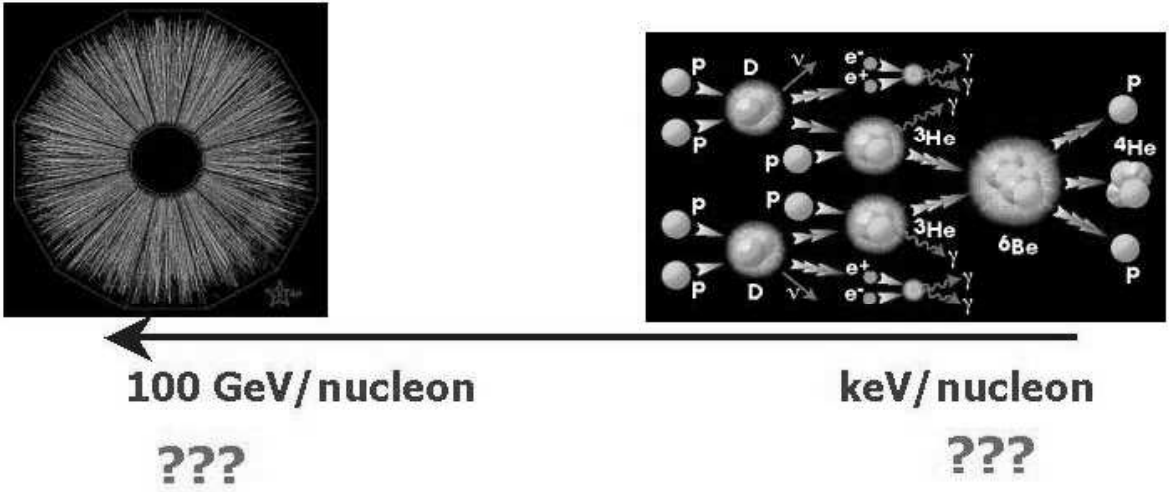}
\end{center}
\caption{{\small The two ends of the energy scale of nuclear reactions with
interest for nuclear astrophysics.}}%
\end{figure}

Nuclear astrophysics at low energies requires the knowledge of the reaction
rate $R_{ij}$ between the nuclei $i$ and $j$. It is given by $R_{ij}%
=n_{i}n_{j}<\sigma v>/(1+\delta_{ij})$, where $\sigma$ is the cross section,
$v$ is the relative velocity between the reaction partners, $n_{i}$ is the
number density of the nuclide $i$, and $<>$ stands for energy average.

In our Sun the reaction $^{7}$Be$\left(  \text{p},\gamma\right)  ^{8}$B plays
a major role for the production of \textit{high energy neutrinos} originated
from the $\beta$-decay of $^{8}B$. These neutrinos come directly from center
of the Sun and are an ideal probe of the Sun's structure. Long ago, Barker
\cite{Bar80} has emphasized that an analysis of the existing experimental data
yields an S-factor for this reaction at low energies which is uncertain by as
much as 30\%. This situation has changed recently, mainly due to the use of
radioactive beam facilities.

The reaction $^{12}C\left(  \alpha,\gamma\right)  ^{16}O$ is extremely
relevant for the fate of massive stars. It determines if the remnant of a
supernova explosion becomes a \textit{black-hole or a neutron star}. It is
argued that the cross section for this reaction should be known to better than
20\%, for a good modelling of the stars \cite{Woo85}. This goal has not yet
been achieved.

\begin{figure}[ptb]
\begin{center}
\includegraphics[
height=3.5in, width=4.5 in ] {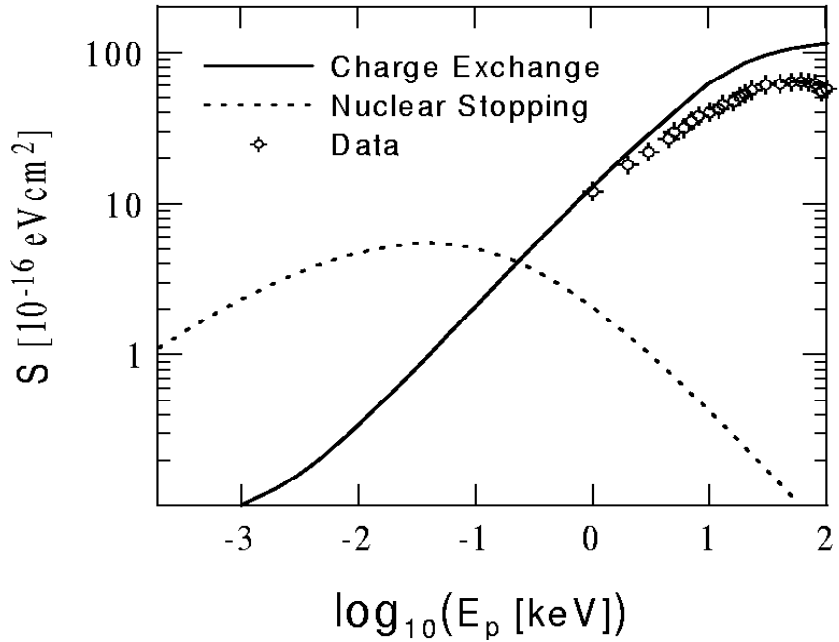}
\end{center}
\caption{{\small The stopping cross section of protons on H-targets. The
dotted line gives the energy transfer by means of nuclear stopping, while the
solid line is the result for the charge-exchange stopping mechanism
\cite{BD00}. The data points are from the tabulation of Andersen and Ziegler
\cite{AZ77}.}}%
\end{figure}

Both the $^{7}$Be$\left(  \text{p},\gamma\right)  ^{8}$B and the $^{12}%
$C$\left(  \alpha,\gamma\right)  ^{16}$O reactions\ cannot be measured at the
energies occurring inside the stars (approximately 20 keV and 300 keV,
respectively). Direct experimental measurements at low energies are often
plagued with low-statistics and large error bars. Extrapolation procedures are
often needed to obtain cross sections in the energy region of astrophysical
relevance. While non-resonant cross sections can be rather well extrapolated
to the low-energy region, the presence of continuum, or subthreshold
resonances, complicates these extrapolations. Numerous radiative capture
reactions pose the same experimental problem.

Approximately half of all stable nuclei observed in nature in the heavy
element region about $A>60$ is produced in the r--process. This r--process
occurs in environments with large neutron densities which lead to$\;\tau
_{\mathrm{n}}\ll\tau_{\beta}$. The most neutron--rich isotopes along the
r--process path have lifetimes of less than one second; typically 10$^{-2}$ to
10$^{-1}$\thinspace s. Cross sections for most of the nuclei involved are hard
to measure experimentally. Sometimes, theoretical calculations of the capture
cross sections as well as the beta--decay half--lives are the only source of
the nuclear physics input for r--process calculations \cite{Sch01}. For nuclei
with about $Z>80$ beta--delayed fission and neutron--induced fission might
also become important.

\section{The Electron Screening Problem}

Besides the Coulomb barrier, nucleosynthesis in stars is complicated by the
presence of electrons. They screen the nuclear charges, therefore increasing
the fusion probability by reducing the Coulomb repulsion. Evidently, the
fusion cross sections measured in the laboratory have to be corrected by the
electron screening when used as inputs of a stellar model. This is a purely
theoretical problem as one can not reproduce the interior of stars in the
laboratory. Applying the Debye-H\"{u}ckel, or Salpeter's, approach
\cite{Sal54}, one finds that the plasma enhances reaction rates, e.g., $^{3}%
$He$(^{3}$He$,\ 2$p$)^{4}$He and $^{7}$Be$($p$,\ \gamma)^{8}$B, by as much as
20\%. This does not account for the dynamic effect due to the motion of the
electrons (see, e.g., \cite{CSK88,Brown97}).

A simpler screening mechanism occurs in laboratory experiments due to the
bound atomic electrons in the nuclear targets. This case has been studied in
great details experimentally, as one can control different charge states of
the projectile+target system in the laboratory
\cite{Ass87,Eng88,Blu90,Rol95,Rol01}. The experimental findings disagree
systematically by a factor of two with theory. This is surprising as the
theory for atomic screening in the laboratory relies on our basic knowledge of
atomic physics. At very low energies one can use the simple adiabatic model in
which the atomic electrons rapidly adjust their orbits to the relative motion
between the nuclei prior to the fusion process. Energy conservation requires
that the larger electronic binding (due to a larger charge of the combined
system) leads to an increase of the relative motion between the nuclei, thus
increasing the fusion cross section. As a matter of fact, this enhancement has
been observed experimentally. The measured values are however not compatible
with the adiabatic estimate \cite{Ass87,Eng88,Blu90,Rol95,Rol01}. Dynamical
calculations have been performed, but they obviously cannot explain the
discrepancy as they include atomic excitations and ionizations which reduce
the energy available for fusion. Other small effects, like vacuum
polarization, atomic and nuclear polarizabilities, relativistic effects, etc.,
have also been considered \cite{BBH97}. But the discrepancy between experiment
and theory remains \cite{BBH97,Rol01}.

\begin{figure}[ptb]
\begin{center}
\includegraphics[
height=3.5in, width=4.2 in ] {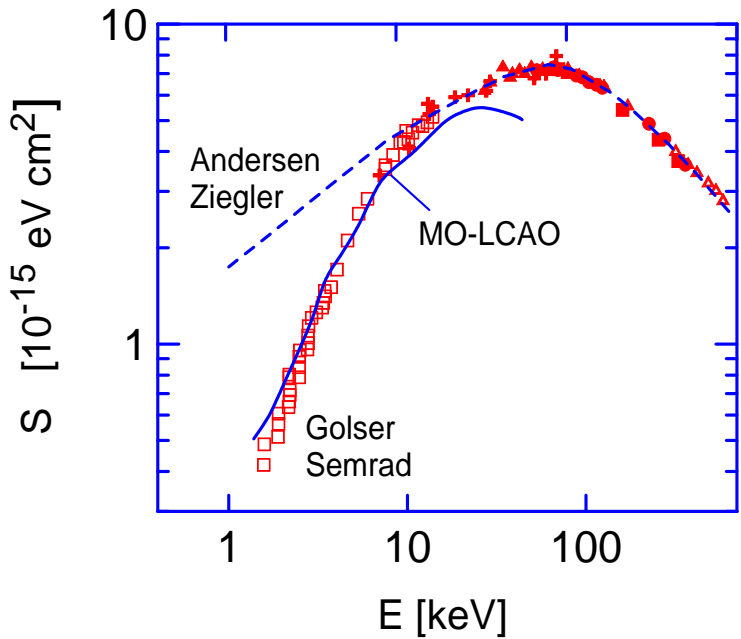}
\end{center}
\caption{{\small The stopping cross section of He$^{+}$ ions on He-targets.
The dashed line is the extrapolation from the Andersen-Ziegler tables. The
solid line is a calculation of ref. \cite{BP03}. The data points are from ref.
\cite{GS91}. The theoretical calculations do not include the nuclear
stopping.}}%
\end{figure}

A possible solution of the laboratory screening problem was proposed in refs.
\cite{LSBR96,BFMH96}. Experimentalists often use the extrapolation of the
Andersen-Ziegler tables \cite{AZ77} to obtain the average value of the
projectile energy due to stopping in the target material. The stopping is due
to ionization, electron-exchange, and other atomic mechanisms. However, the
extrapolation is challenged by theoretical calculations which predict a lower
stopping. Smaller stopping was indeed verified experimentally \cite{Rol01}. At
very low energies, it is thought that the stopping mechanism is mainly due to
electron exchange between projectile and target. This has been studied in ref.
\cite{BD00} in the simplest situation; proton+hydrogen collisions (see figure
2). Two-center electronic orbitals were used as input of a coupled-channels
calculation. The final occupation amplitudes were projected onto bound-states
in the target and in the projectile. The calculated stopping power was added
to the nuclear stopping power mechanism, i.e. to the energy loss by the
Coulomb repulsion between the nuclei. The obtained stopping power is
proportional to $v^{\alpha}$, where $v$ is the projectile velocity and
$\alpha=1.35$. The extrapolations from the Andersen-Ziegler table predict a
larger value of $\alpha$. Although this result seems to indicate the stopping
mechanism as a possible reason for the laboratory screening problem, the
theoretical calculations tend to disagree on the power of v at low energy
collisions. For example, ref. \cite{GS91} found $S\sim v_{p}^{3.34}$ for
protons in the energy range of 4 keV incident on helium targets. This is an
even larger deviation from the extrapolations of the Andersen-Ziegler tables.

\begin{figure}[ptb]
\begin{center}
\includegraphics[
height=3. in, width=3.5 in ] {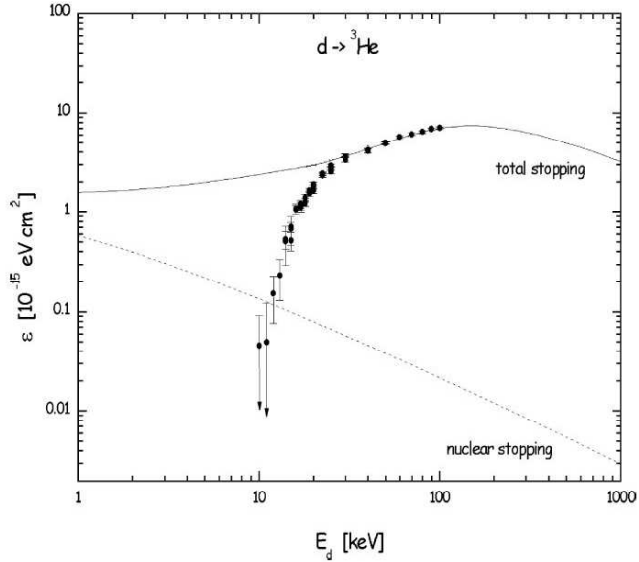}
\end{center}
\caption{{\small Energy loss of deuterons in }$^{3}${\small He gas as a
function of the deuteron energy. Data are form ref. \cite{Form}.}}%
\end{figure}

Recently, we have performed \cite{BP03} a calculation of the stopping power in
atomic He$^{+}+$He collisions using the two-center molecular orbital basis
(MO). The advantage of the use of (MO) basis over the method of ref. \cite{GS}
is that the numerical calculation converges more rapidly at small nuclear
distances. The disadvantage is that it converges slower at larger nuclear
distances. However this can be easily corrected by using a linear combination
of atomic orbitals (LCAO) at large distances. A much smaller basis than in
ref. \cite{GS} is needed in both situations. One also has to account for the
level crossing problem in the adiabatic collision model. This has been done in
the Landau-Zenner approximation \cite{Lan32}. It has been shown that the level
crossing is practically diabatic, i.e., as if no level repulsion were present
\cite{BP03}. The result of this calculation is shown in figure 3. The
agreement with the data from ref. \cite{GS91} at low energies is excellent.
However, the theoretical calculations do not include the nuclear recoil. The
agreement with the data disappears completely if the nuclear recoil is
included. In fact, the unexpected "disappearance" of the nuclear recoil was
also observed in ref. \cite{Form} (see figure 4). This seems to violate a
basic principle of nature, as the nuclear recoil is due to the Coulomb
repulsion between the projectile and the target atoms \cite{AZ77}. This effect
should always be present in the experimental data.

\begin{figure}[ptb]
\begin{center}
\includegraphics[
height=3.in, width=3.5 in ] {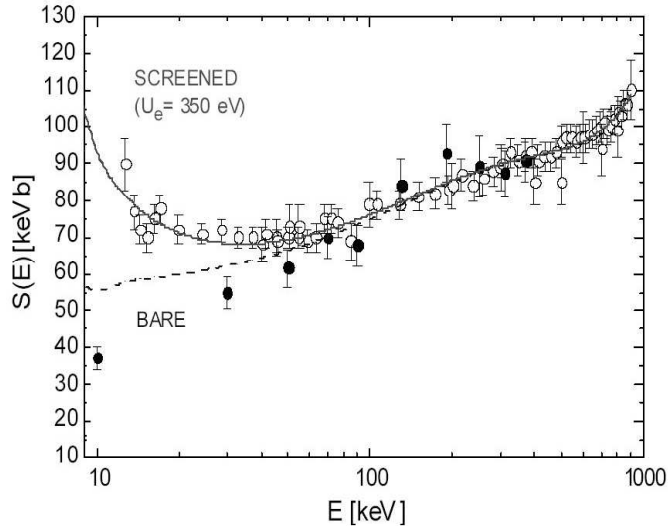}
\end{center}
\caption{{\small S-factor for the reaction }$^{7}$Li$($p$,\alpha)\alpha
${\small \ obtained by the measurement of the cross section for the reaction
}$^{2}$He$($Li$,\alpha\alpha)$n{\small \ with the analysis based on the
Trojan-Horse method \cite{Al00}. The solid curve is a theoretical fit to the
direct measurement data for }$^{7}$Li$($p$,\alpha)\alpha$ assuming a screening
energy of 350 eV. The dashed curve is the extrapolation of the S-factor to low
energies, assuming no screening effect. }%
\end{figure}

We are faced here with a notorious case of obscurity in nuclear astrophysics.
The disturbing conclusion is that as long as we cannot understand the
magnitude of electron screening in stars or in the atomic electrons in the
laboratory, it will be even more difficult to understand color screening in a
quark-gluon plasma, an important tool in relativistic heavy ion physics (e.g.,
the $J/\Psi$ suppression mechanism) \cite{MS86}.

\section{Radioactive Beam Facilities and Indirect Methods}

Transfer reactions are a well established tool to obtain spin, parities,
energy, and spectroscopic factors of states in a nuclear system.
Experimentally, $(d,\ p)$ reactions are mostly used due to the simplicity of
the deuteron. Variations of this method have been proposed by several authors.
For example, the \textit{Trojan Horse} Method was proposed in ref.
\cite{Bau86} (see also \cite{TP03}) as a way to overcome the Coulomb barrier.
If the Fermi momentum of the particle $x$ inside $a=(b+x)$ compensates for the
initial projectile velocity $v_{a}$, the low energy reaction $A+x=B+c$ is
induced at very low (even vanishing) relative energy between $A$ and $x$.
Successful applications of this method has been reported recently
\cite{Spit00}. Figure 5 shows the astrophysical S-factor for the reaction
$^{7}\mathrm{Li}({\normalsize p},\alpha)\alpha$ obtained by the measurement of
the cross section for the reaction $^{2}\mathrm{He}(\mathrm{Li},\alpha
\alpha)\mathrm{n}$ with the analysis based on the Trojan-Horse method. One
sees that the problems related to the atomic electron screening are not
present in the experimental data. This clearly shows, for the first time, the
advantage of using this technique over the direct measurements, as one avoids
the treatment of the mysterious screening problem.

Recently the \textit{stripping reactions} have been demonstrated to be a
useful tool to deduce spectroscopic factors in many reactions of relevance for
nuclear astrophysics \cite{HJ03}. The method is also free of the screening problem.

\begin{figure}[ptb]
\begin{center}
\includegraphics[
height=3.in, width=3.5 in ] {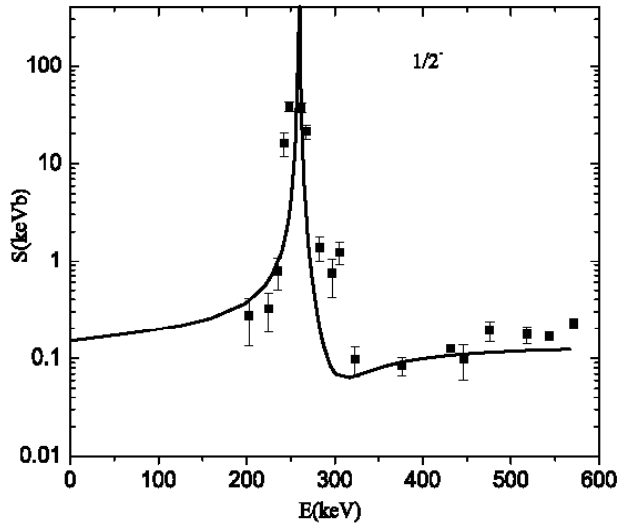}
\end{center}
\caption{{\small The astophysical S-factor for the reaction }$^{13}%
N(p,\gamma)^{15}O${\small \ obtained by using the asymptotic normalization
coefficient technique \cite{Mu03}. Only the S-factor for the capture to the
1/2}$^{-}${\small \ ground state is shown. The techinque also allows for the
measurement of S-factors to excited states.}}%
\end{figure}

At low energies the amplitude for the radiative capture cross section is
dominated by contributions from large relative distances of the participating
nuclei. Thus, what matters for the calculation of the direct capture matrix
elements are the \textit{asymptotic normalization coefficients} (ANC). This
coefficient is the product of the spectroscopic factor and a normalization
constant which depends on the details of the wave function in the interior
part of the potential. The normalization coefficients can be found from
peripheral transfer reactions whose amplitudes contain the same overlap
function as the amplitude of the corresponding astrophysical radiative capture
cross section. This idea was proposed in ref. \cite{Muk90} and many successful
applications of the method have been obtained \cite{Cag}. For example, the
astrophysical S-factor for the reaction $^{13}N(p,\gamma)^{15}O$ obtained by
using the asymptotic normalization coefficient technique \cite{Mu03} (see
figure 6). Only the S-factor for the capture to the 1/2$^{-}$ ground state is
shown. The technique also allows for the measurement of S-factors to excited states.

\textit{Charge exchange} induced in $(p,n)$ reactions are often used to obtain
values of Gamow-Teller matrix elements which cannot be extracted from
beta-decay experiments. This approach relies on the similarity in spin-isospin
space of charge-exchange reactions and $\beta$-decay operators. As a result of
this similarity, the cross section $\sigma($p,\ n$)$ at small momentum
transfer $q$ is closely proportional to $B(GT)$ for strong transitions
\cite{Tad87}. As shown in ref. \cite{Aus94}, for important GT transitions
whose strength are a small fraction of the sum rule the direct relationship
between $\sigma($p,\ n$)$ and $B(GT)$ values fails to exist. Similar
discrepancies have been observed \cite{Wat85} for reactions on some odd-A
nuclei including $^{13}$C, $^{15}$N, $^{35}$Cl, and $^{39}$K and for
charge-exchange induced by heavy ions \cite{Ber96,St96}.

\begin{figure}[ptb]
\begin{center}
\includegraphics[
height=3.in, width=3.5 in ] {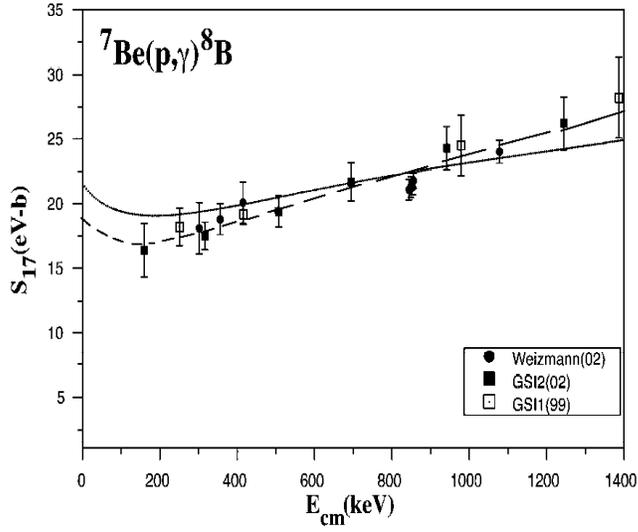}
\end{center}
\caption{{\small The astrophysical S-factor for the reaction }$^{7}$%
Be$($p$,\gamma)^{8}$B{\small \ at low energies. The data points from direct
measurements (Weizmann) are from ref. \cite{bab03}. The GSI data were obtained
by using the Coulomb dissociation method \cite{Iwa99,Sue03}. The solid curve
is a calculation from ref. \cite{Des94} and the dashed curve is a calculation
from refs. \cite{Ber94,Typ02}. }}%
\end{figure}

The (differential, or angle integrated) \textit{Coulomb breakup} cross section
for $a+A\longrightarrow b+x+A$ can be written as $\sigma_{C}^{\pi\lambda
}(\omega)=F^{\pi\lambda}(\omega)\ .\ \sigma_{\gamma}^{\pi\lambda}(\omega)$,
where $\omega$ is the energy transferred from the relative motion to the
breakup, and $\sigma_{\gamma}^{\pi\lambda}(\omega)$ is the photo nuclear cross
section for the multipolarity ${\pi\lambda}$ and photon energy $\omega$. The
function $F^{\pi\lambda}$ depends on $\omega$, the relative motion energy, and
nuclear charges and radii. They can be easily calculated \cite{Ber88} for each
multipolarity ${\pi\lambda}$. Time reversal allows one to deduce the radiative
capture cross section $b+x\longrightarrow a+\gamma$ from $\sigma_{\gamma}%
^{\pi\lambda}(\omega)$. This method was proposed in ref. \cite{BBR86}. It has
been tested successfully in a number of reactions of interest for astrophysics
(\cite{Bau94} and references therein). The most celebrated case is the
reaction $^{7}$Be$($p$,\gamma)^{8}$B. It has been studied in numerous
experiments in the last decade. For a recent compilation of the results
obtained with the method, see e.g. ref. \cite{DT03} (see also figure 7). They
have obtained an $S_{17}(0)$ value of 19.0 eV.b which is compatible with the
value commonly used in solar model calculations \cite{Bah89}. To achieve the
goal of applying this method to many other radiative capture reactions (for a
list, see, e.g. \cite{Bau94}), detailed studies of dynamic contributions to
the breakup have to be performed, as shown in refs. \cite{BBK92,BB93}. The
role of higher multipolarities (e.g., E2 contributions \cite{Ber94,GB95,EB96}
in the reaction $^{7}$Be$($p$,\gamma)^{8}$B) and the coupling to high-lying
states \cite{Ber02} has also to be investigated carefully. In the later case,
a recent work has shown that the influence of giant resonance states is small
(see figure 8). Studies of the role of the nuclear interaction in the breakup
process is also essential to determine if the Coulomb dissociation method is
useful for a given system \cite{BD03}.

\begin{figure}[ptb]
\begin{center}
\includegraphics[
height=3.in, width=3.5 in ] {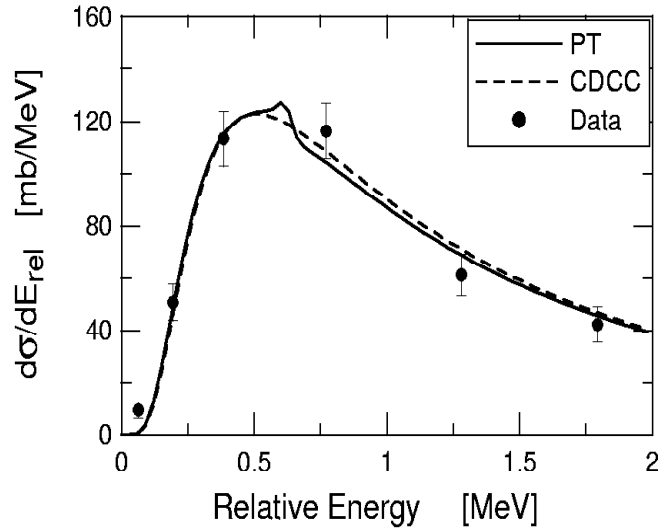}
\end{center}
\caption{{\small Energy dependence of the Coulomb breakup cross section for
$^{8}\mathrm{B}+\mathrm{Pb}\longrightarrow\mathrm{p}+^{7}\mathrm{Be}%
+\mathrm{Pb}$ at 84 MeV/nucleon. First-order perturbation calculations (PT)
are shown by the solid curve. The dashed curve is the result of a CDCC
calculation including the coupling between the ground state and the low-lying
states with the giant dipole and quadrupole resonances \cite{Ber02}. The data
points are from ref. \cite{Da01}.}}%
\end{figure}

In summary, radioactive beam facilities have opened a new paved way to
disclosure many unknown features of reactions in stars and elsewhere in the universe.

\ \

Work Supported by U.S. National Science Foundation under Grants No. PHY-007091
and PHY-00-70818.

\end{document}